\documentclass[sigconf, authorversion=true, nonacm=true]{acmart}  
\AtBeginDocument{%
  \providecommand\BibTeX{{%
    \normalfont B\kern-0.5em{\scshape i\kern-0.25em b}\kern-0.8em\TeX}}}

\usepackage[utf8]{inputenc} 
\usepackage[T1]{fontenc}    
\usepackage{hyperref}       
\usepackage{url}            
\usepackage{booktabs}       
\usepackage{amsfonts}       
\usepackage{nicefrac}       
\usepackage{microtype}      
\usepackage{xcolor}         
\usepackage{subfig}
\usepackage{float}
\usepackage{graphicx}
\usepackage{amsmath}
\usepackage{enumitem}

\newcommand{\statcheck}[1]{\textcolor{black}{#1}}
\newcommand{\ours}[1]{\textcolor{black}{adaptive personalization}}

\setitemize{label=\textbullet, leftmargin=*, nolistsep}

\copyrightyear{2024} 
\acmYear{2024} 
\setcopyright{acmlicensed}
\acmConference[HRI '24]{Proceedings of the 2024 ACM/IEEE International Conference on Human-Robot Interaction}{March 11--14, 2024}{Boulder, CO, USA}
\acmBooktitle{Proceedings of the 2024 ACM/IEEE International Conference on Human-Robot Interaction (HRI '24), March 11--14, 2024, Boulder, CO, USA}
\acmPrice{15.00}
\acmDOI{10.1145/3610977.3635000}
\acmISBN{979-8-4007-0322-5/24/03}

\begin{document}

\title{Towards Balancing Preference and Performance through Adaptive Personalized Explainability}


\author{Andrew Silva}
\authornote{Work completed while at the Georgia Institute of Technology. This work reflects solely the opinions and conclusions of the authors and not of TRI or any Toyota entity.}
\email{andrew.silva@tri.global}
\orcid{0000-0002-0317-5135}
\affiliation{
\institution{Toyota Research Institute}
\city{Cambridge}
\state{Massachusetts}
\country{USA}
}

\author{Pradyumna Tambwekar}
\email{ptambwekar3@gatech.edu}
\orcid{}
\affiliation{
\institution{Georgia Institute of Technology}
\city{Atlanta}
\state{Georgia}
\country{USA}
}

\author{Mariah Schrum}
\authornotemark[1]
\email{mariahschrum@berkeley.edu}
\orcid{}
\affiliation{
\institution{University of California, Berkeley}
\city{Berkeley}
\state{California}
\country{USA}
}

\author{Matthew Gombolay}
\email{matthew.gombolay@cc.gatech.edu}
\orcid{}
\affiliation{
\institution{Georgia Institute of Technology}
\city{Atlanta}
\state{Georgia}
\country{USA}
}

\begin{abstract}
  As robots and digital assistants are deployed in the real world, these agents must be able to communicate their decision-making criteria to build trust, improve human-robot teaming, and enable collaboration. While the field of explainable artificial intelligence (xAI) has made great strides to enable such communication, these advances often assume that one xAI approach is ideally suited to each problem (e.g., decision trees to explain how to triage patients in an emergency or feature-importance maps to explain radiology reports). This fails to recognize that users have diverse experiences or preferences for interaction modalities. In this work, we present two user-studies set in a simulated autonomous vehicle (AV) domain. We investigate (1) population-level preferences for xAI and (2) personalization strategies for providing robot explanations. We find significant differences between xAI modes (language explanations, feature-importance maps, and decision trees) in both preference ($p < 0.01$) and performance ($p < 0.05$). We also observe that a participant's preferences do not always align with their performance, motivating our development of an adaptive personalization strategy to balance the two. We show that this strategy yields significant performance gains ($p < 0.05$), and we conclude with a discussion of our findings and implications for xAI in human-robot interactions.
\end{abstract}

\ccsdesc[500]{Human-centered computing~Human computer interaction (HCI)}
\ccsdesc[500]{Computing methodologies~Artificial intelligence}

\keywords{Explainability, Personalization, User Studies}



\maketitle

\section{Introduction}
As robots and digital assistants are deployed to the real world, these agents must be able to communicate their decision-making criteria to build trust, improve human-robot teaming, and enable collaboration \cite{Communication,Paleja2021Minecraft}. Researchers have identified \textit{explainability} as a necessary component of high-quality human-robot interactions in many domains \cite{doshi2017towards,rudin2021interpretable}. While several approaches for explainability are under active investigation (e.g., natural language explanations \cite{deyoung2019eraser}, decision-tree extraction \cite{pmlr-v108-silva20a}, counterfactual presentation \cite{karimi2021algorithmic}, saliency-based explanations \cite{suau2020finding,lime}, etc.), existing studies on human-use of explanations is almost entirely confined to treating explanation as a ``one-size-fits-all'' problem \cite{madumal2020explainable,poursabzi2021manipulating,mullenbach2018explainable}. However, explanations have different functional roles with respect to deployment context \cite{arya2019one, gilpin2022explanation}, suggesting that personalization and contextualization of explanations is an important and understudied avenue to bring explainability to the real world.
If individual preferences and expertise affect the success of an explanation~\cite{tambwekar2023towards}, a natural next step is to identify which xAI modalities should be shown to an individual user for any given decision. 

Within the field of xAI, simply measuring the accuracy or fidelity of an explanation (with respect to the underlying agent or algorithm) is not enough to know that an explanation was useful.
If explanations do not carefully consider a user's expertise or expectations, the simple act of showing an explanation can cause the user to blindly trust an agent's advice, leading to adverse effects on performance and trust \cite{poursabzi2021manipulating,silva2022explainable}.
This counter-intuitive result presents a key problem: explanations encourage inappropriate compliance. If users see explanations and defer to robots without critically examining the robot suggestion, then researchers must develop a deeper understanding of the relationship between explanations and compliance while also improving an xAI agent's ability to expose faulty decision-making to human users~\cite{ehsan2021explainability}. By improving our understanding of such relationships and by better calibrating to end-users, we can produce xAI systems that are not only easier and more enjoyable to use, but also improve outcomes and efficiency of human users \cite{ schlicker2021towards, ferrario2022explainability}.

Ultimately, xAI research seeks to help humans understand when to rely on vs. override their AI assistants, using explanations to determine if decisions are sound and trustworthy~\cite{hoffman2018metrics}. Such a dynamic exists when humans collaborate with fallible AI assistants-- a scenario that we recreate in this work. Our work aims to understand the diverse preferences of untrained humans with potentially-faulty assistants that use xAI to support human decision making. We present a set of studies in which participants interact with a virtual AV to navigate through an unknown city with the assistance of a digital agent, replicating a common problem of navigating in a new place. Crucially, this assistive agent is not always correct, and incorrect advice is signalled with the inclusion of red-herring features (e.g., if the agent refers to ``weather'' in its explanation, the suggestion is wrong). Our work therefore simulates the use of xAI for explanatory debugging \cite{das2021xai,das2023subgoals} with concept-based explanations \cite{das2023state2explanation}, also called the ``glitch detector task'' \cite{hoffman2001storm,hoffman2023measures}. We investigate how xAI may improve people's mental models for AI \cite{anderson2020mental, brachman2023follow}, and how \textit{personalized} xAI will affect people's ability to accurately identify when their assistant is correct or incorrect (i.e., if the agent adapts to the user, will the user make fewer mistakes?). Our contributions include:
\begin{enumerate}
    \item We design two studies in which participants interact with xAI modalities randomly or using personalization. 
    \item We empirically study participants' preferences for certain types of explanations, as well as their performance with such explanations, finding that language explanations are significantly preferred ($p < 0.05$) and lead to fewer mistakes ($p < 0.05$) relative to other modalities.
    \item We develop a novel adaptive personalization approach to dynamically balance a participant's preference- and performance-based needs depending on their progress in a task.
    \item  We find that adapting to a participant's preferences while also maximizing their performance leads to fewer mistakes relative to naive, randomly-chosen explanations or a preference-maximization approach ($p < 0.05$), and leads to significantly greater perceptions of preference-accommodation relative to an agent that does not personalize ($p < 0.05$). 
\end{enumerate}

Unlike prior work on personalizing xAI, which only considers user-preferences~\cite{conati2021toward, lai2023selective, li2023personalized, kouki2020generating}, our work is the first to directly account for the user's task-performance when personalizing xAI. Our work takes a crucial first step towards understanding \textit{how} future work should consider personalization in xAI as well as \textit{why} such personalization matters.

\section{Related work}
\label{sec:related-work}
In this work, we investigate the effects of personalizing xAI mechanisms to users, focusing on three primary modalities for explanation: language generation and counterfactuals \cite{karimi2020survey,karimi2021algorithmic,2023hri_language}, feature-importance maps \cite{lime,2023visuotextual}, and decision-tree explanations \cite{pmlr-v108-silva20a,rudin2021interpretable}. 
We investigate the domain of AVs to study personalized xAI, a domain of increasing interest to the HRI community~\cite{fink2023, abrams2021, Li2020, moore2020}.

\textbf{Personalization -- }
With the proliferation of digital assistants and machine learning in consumer products, the problem of personalization has become more pressing. While conventional machine learning applies a single model to all data, the real-world contains many problems where the same sample may have different labels depending on the user (e.g., people wanting to customize a social robot's greeting, behavior, or appearance \cite{2023_personal_neuro,2023hri_social_personalization,2021hri_personalized_wigs}). This problem setup requires personalization of the shared model, such that individual users can receive personally-tailored responses from the learned model \cite{collins2021fedrep,li2019fedmd}, ideally without needing to retrain the entire model from scratch.
There have been several approaches for personalization with such shared models, including personal model heads for each user \cite{arivazhagan2019federated,collins2021fedrep,dinh2020pfedme,kim2021spatio,li2019fedmd,li2020federated,paulik2021federated,rudovic2021personalized} or meta-learned models that can rapidly adapt to users \cite{chen2022fast,deng2020apfl,fallah2020personalized,hanzely2020lower,hanzely2020federated,jiang2019improving,lin2019personalizing}. 

\textbf{Personal Embeddings --} In this work, we build on personalization via personal embeddings \cite{hsiao2019learning,paleja2020interpretable,schrum2022reciprocal,schrum2022mind,silva2022fedembed,fedperc,tamar2018imitation}, in which a unique embedding is assigned to each user and appended to the input data or hidden representations of the network, thereby allowing the model to adapt its decision-making by conditioning on this unique per-person embedding. 

\textbf{Adaptive Personalization --}  While personalizing to different human users in this work, our system must balance between two objectives-- user-preference and task-performance. We refer to this balancing act as ``adaptivity''. Prior work \cite{axelsson-skantze-2019-modelling,axelsson2023,robrecht2023study} also develops ``adaptive'' approaches to personalization, though prior definitions only consider task-performance in a single domain. In contrast, our approach is generally applicable to any explanatory debugging scenario with concept-based explanations, and dynamically balances between task-performance and human preferences.

\textbf{Explainability -- }
While personalization helps to bring machine learning to wider audiences and a greater diversity of problems, the inherent unpredictability of models remains an obstacle to wider deployment of learned solutions. There are legal \cite{voigt2017eu} and practical criteria for machine learning models to be used in many contexts \cite{doshi2017towards,doshi-velez_accountability_2017}.
xAI is a subfield of machine learning research seeking to help justify a model's decision-making using a variety of approaches. The field has contributed many techniques, such as developing neural network models that can be readily interpreted \cite{wang2016bayesian}, pursuing natural language generation for explanations \cite{chen2021generate, ehsan2019automated,mullenbach2018explainable,wiegreffe-pinter-2019-attention}, presenting feature-importance maps for input samples \cite{gaams,jain-wallace-2019-attention,lime,vstrumbelj2014explaining,suau2020finding,wiegreffe-pinter-2019-attention,2023visuotextual}, presenting relevant training data \cite{brunet2018understanding,caruana1999case,koh2017understanding,clif}, and other techniques \cite{hoffman2018metrics,holzinger2020measuring,linardatos2021explainable}.

\begin{figure*}[t]\centerline{\includegraphics[width=0.85\textwidth]{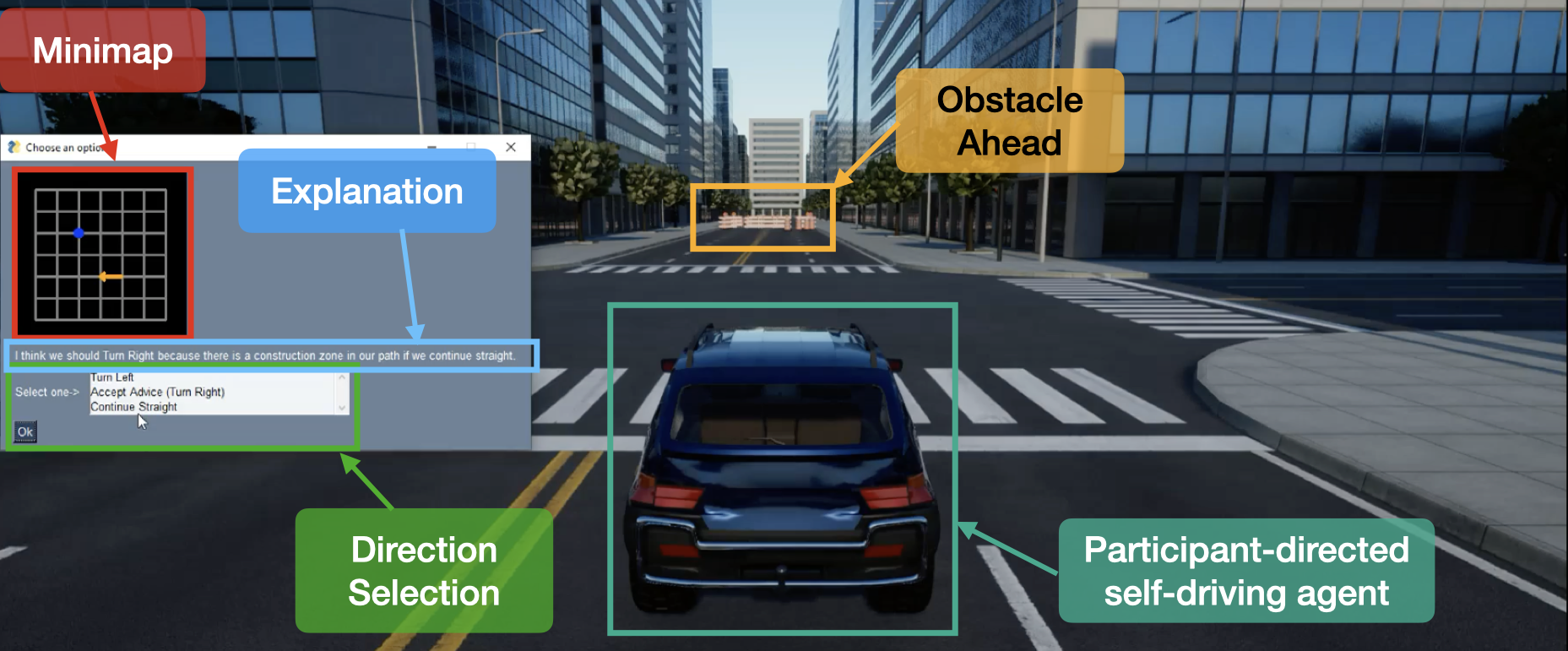}}
\caption{Here we show an example interaction with a language explanation and a correct suggestion. Taking a wrong turn (e.g., going straight) will lead directly into a roadblock, forcing participants return to this intersection and repeat the interaction.}
\label{fig:interface}
\end{figure*}

While research has begun to investigate the effects of explanations on user's ability to understand and forecast network behavior \cite{anderson2020mental,hase-bansal-2020-evaluating,hoffman2018metrics,hutton2012crowdsourcing,madumal2020explainable,nguyen-2018-comparing,poursabzi2021manipulating,pmlr-v108-silva20a,silva2022explainable,sokol2020explainability,tintarev2012evaluating} or the social implications of working with robots that explain their behavior \cite{2023_hri_clinical_xai,2023_hri_xai_repairs_trust,devleena_xai,2020hri_xai_acceptance}, there is considerably less work on understanding how such dynamics would unfold if a human were afforded the ability to influence the xAI agent more directly, such as through controlling what types of explanations are provided. Prior work has studied the effects of explanations on compliance with inaccurate suggestions \cite{poursabzi2021manipulating,silva2022explainable}; however, it is possible that such explanations were simply ill-suited to the study participants and that personalization would help mitigate this problem. 
Prior work has also shown that user-characteristics and dispositional factors can significantly impact a user's interaction with an explainable agent~\cite{millecamp2019explain, millecamp2020s, shulner2022enhancing, shulner2022fairness}, and that aligning explanations with a user's expertise can lead to higher perceived utility \cite{park2022perception}.
In this work, we address these oversights in prior work and enable active personalization based on user feedback, studying compliance with personalized xAI.

\section{Study Setup}
\label{sec:study-setup}
In our work, we present two separate studies using the same environment and driving agent. The first is a \textbf{population study} with a within-subjects design targeted at identifying population-wide trends, and serving as a data capture for our personalization model. The second is a \textbf{personalization study} with a mixed design to test the effects of different personalization strategies on how well they align with participant's preferences and how effectively they maximize task performance. We also present the design of an adaptive personalization agent that seeks to jointly satisfy both objectives. In this section, we provide background for the shared environment, driving agent, metrics, and research questions in the two studies, before detailing study procedures, metrics, and results in Sections \ref{sec:phase1} \& \ref{sec:phase2}. All studies in this work took 60-75 minutes, and participants were compensated \$20 for their time. All studies were approved by an Institutional Review Board (IRB).
\begin{figure*}[t]\centerline{\includegraphics[width=\textwidth]{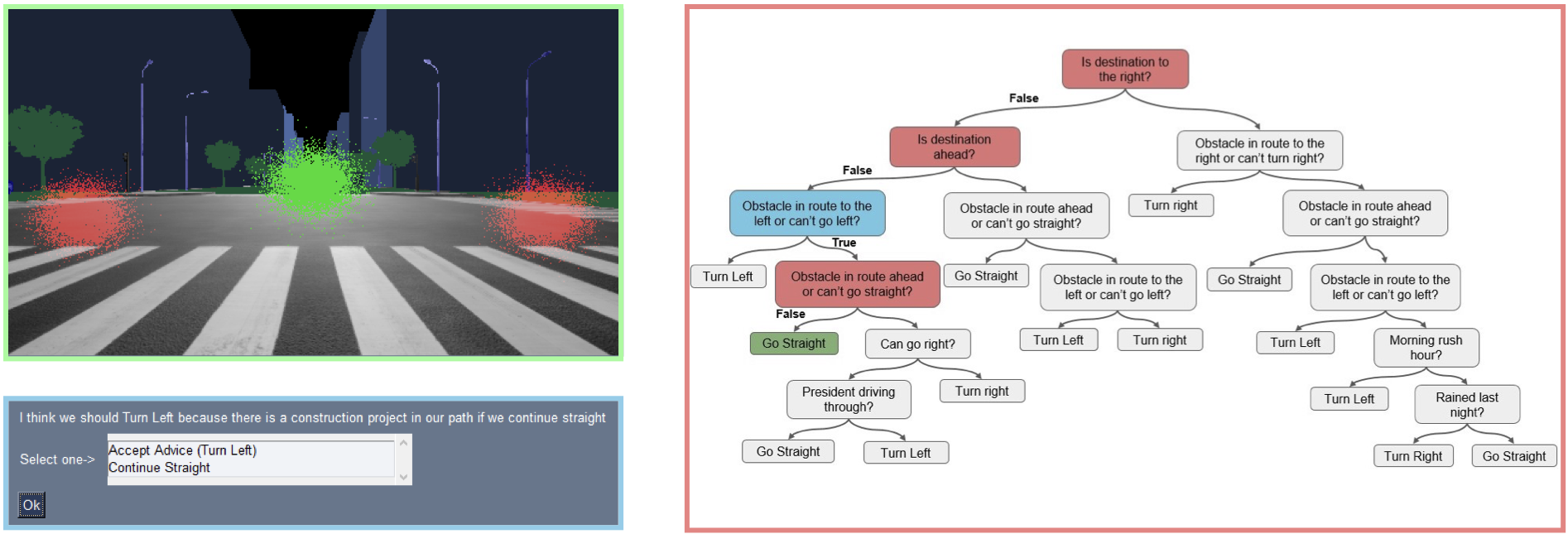}}
\caption{We compare three xAI modalities in this work: feature-importance maps, (top left) in which highlighted regions indicate possible directions and relevant elements of the image, such as green indicating the suggested direction, language explanations (bottom left) that are a sentence justifying one direction over another, and decision trees (right) in which the highlighted path leads to the suggested direction. Red blocks mean ``false'' and blue blocks mean ``true''.}
\label{fig:explanations}

\end{figure*}

\vspace{-0.1cm}
 \subsection{Environment}
 \label{subsec:domain}
The domain we employed for our experiments was a simulated driving domain. The participant interacts with an AV, reflecting a robot-deployment that is becoming increasingly common in the real world.
Furthermore, autonomous driving is an accessible and easily-understandable domain for a non-expert end-user, and is a highly pertinent area of study for human-robot communication and xAI \cite{2023xai_av1,2023xai_av2}. 
In our study, the human is responsible for all navigational direction, but the robot handles all actual control of the vehicle. This task was setup through the AirSim driving simulator~\cite{shah2018airsim} and built in the Unreal Engine. 
Our domain features a simulated city with a seven-by-seven grid layout, effectively putting the participants into a small maze that they will navigate for the duration of each task. 
Participants direct an AV through the maze to the goal location in the city, working with assistance from a self-driving agent and a small mini-map.

Each intersection in the domain presents an opportunity to select a direction to progress through the environment, giving the participant all available navigation options (e.g. ``turn left,'' ``turn right,'' or ``continue straight'') alongside a directional suggestion from the self-driving agent and an explanation justifying the suggestion. Participants consider the suggestions and explanations to help them decide how to navigate through the city. After making a decision, participants are also asked to provide binary positive/negative feedback on whether or not they would like to see more explanations with the modality that they received. We provide an example of one such interaction in Figure \ref{fig:interface}. 

The city contains several roadblocks, thereby creating a single optimal path to the goal, with any deviation resulting in either a U-turn (as the car drives down a road with a roadblock and must turn around) or a significantly slower path.
For each task, the participant has a new starting and goal location, and roadblocks are moved around the map. This relocation prevents participants from memorizing routes through the city, and encourages reliance on navigational assistance from the self-driving agent.

 \subsection{AI Driving Agent and Explanations}
 \label{subsec:driving-agent}
At each intersection in the domain, a digital agent suggests a direction and also presents an explanation for its suggestion to the participant. Explanations are either a sentence in natural language, a feature-importance map, or a decision-tree (Figure \ref{fig:explanations}). All explanations were manually generated before the study, rather than autonomously generated via an existing machine learning method \cite{pmlr-v108-silva20a,karimi2021algorithmic,lime}, to control for existing explainability research and to more closely examine the modalities themselves. To identify the optimal direction, the agent uses a breadth-first search planner over the grid to find the shortest path to the goal.

Approximately 30\% of the time, the agent will suggest the \textit{opposite} of the optimal direction, alongside a flawed explanation attempting to rationalize the incorrect suggestion.  Participants are trained to identify these \textit{incorrect} explanations before beginning the study. The threshold for performance was chosen following prior work on agent reliability in user studies \cite{wiczorek2014supporting,yang2017evaluating,parasuraman2010}. Further details on how incorrect suggestions are provided and signalled are given in the supplementary material.

\subsection{Metrics}
\label{subsec:shared-metrics}

\textbf{Shared Metrics --} In both studies, we employ the following metrics: 
\begin{itemize}
    \item \textit{Inappropriate Compliance} - The proportion of incorrect advice accepted by participants. The better a participant understands a particular xAI method, the lower this metric should be.
    \item \textit{Mistakes} - The number of mistakes made by the participant, as an additional gauge of the participant's ability to interpret an explanation modality.
    \item \textit{Binary Feedback Ratings} - Participants' answers to a "yes/no" question about whether the participant would like to work with a specific xAI modality again, which is asked to the participant after every intersection. We record the total number of positive and negative responses for each xAI modality across the study.
\end{itemize}

\vspace{0.1cm}

\noindent \textbf{Population Study Metrics --} In the population study, we also measure:
\begin{itemize}
    \item \textit{Preference Rankings} - Rankings from a 5-item ranking survey between the explanation modalities. These values will be high if participants felt that the condition did a better job of accommodating their personal preferences. 
    \item \textit{Consecutive mistakes} - Back-to-back mistakes as a result of rejecting correct suggestions or accepting incorrect suggestions. 
    \item \textit{Consideration Time} - The amount of time a participant considers an explanation prior to making a decision. 
\end{itemize}
In the population study, participants work with a fixed xAI modality for an entire task. Therefore, through measuring consecutive mistakes, we are able to infer a participants' reaction to making a mistake, i.e. does the specific xAI modality enable them to better reflect on what they missed in the previous iteration, or will they make repeat mistakes? Similarly, consideration time tells us if one modality is slower or faster than others.

\vspace{0.1cm}
\noindent \textbf{Personalization Study Metrics --} Finally, the personalization study also measures:
\begin{itemize}
    \item \textit{Steps Above Optimal} - The number of steps to complete a task, relative to the optimal solution.
    \item \textit{Preference Annotations} - Free form text responses to how well the different agents accommodated participant preferences. Free form text allowed participants to describe the various successes and failures of different personalization strategies without being confined to a predefined ranking survey.
\end{itemize}
We do not measure consideration time or consecutive mistakes in this study, as such metrics target the xAI modalities themselves rather than the personalization strategies we seek to compare and because the xAI modality can change between interactions (i.e., modalities are not fixed, as in the population study). We also change the approach to measuring preference, giving us more insight into why participants preferred one option over another by requiring participants to describe their preferences \cite{2023hri_social_personalization}.

\subsection{Research Questions}
Our work aims to understand both (1) population-wide trends on preference and performance for diverse xAI modalities, and (2) the effects of different personalization strategies on human-robot teaming with xAI. As our study uses a novel domain, we first sought to verify whether there was a specific modality that led to the highest average preference or performance. Furthermore, recent work has highlighted a nuanced relationship between preference and performance, often in relation to external factors, such as expertise~\cite{Paleja2021Minecraft, tambwekar2023towards}. To gain insight into these relationships, our population study is designed with the following research questions in mind:
\begin{itemize}
    \item \textbf{RQ1.1 -- Preferences}: Will one xAI modality be significantly more preferred than others? 
    \item \textbf{RQ1.2 -- Performance}: Will one xAI modality lead to significantly better performance than others?
    \item \textbf{RQ1.3 -- Alignment}: Will participants prefer to use the modality that maximizes their performance?
\end{itemize}

The personalization study seeks to examine the degree to which balanced personalization affects participants' performance on a task and on their perceptions of the agent's accommodation of their preferences (i.e., does balanced personalization make people feel like the agent is listening to them while also helping them perform better?). The primary research questions are then:
\begin{itemize}
    \item \textbf{RQ 2.1 -- Preferences} Will balanced personalization be significantly more preferred than other personalization strategies?
    \item \textbf{RQ 2.2 -- Performance} Will balanced personalization lead to significantly fewer mistakes than other personalization strategies?
    \item \textbf{RQ 2.3 -- Comparison to known-best} Will balanced personalization match or exceed task-performance and preference metrics when compared to the a-priori known best xAI modality for the study task (i.e., language explanations). 
\end{itemize}

\section{Population Study}
\label{sec:phase1}
The population study enables us to study overall trends for preference and task-performance with our chosen xAI modalities and domain. This study helps to determine which mode, if any, is superior for this task and enables us to clearly analyze the relationship between performance and task-preference for each xAI modality.

\subsection{Study Conditions}
\label{subsec:phase1-conditions}
The population study is a within-subjects design to study the effects of xAI modalities. Therefore, the conditions in this study are the xAI modalities themselves (Section \ref{subsec:driving-agent}), including (1) Language, (2) Feature Maps, and (3) Decision Tree explanations.
Each of these conditions were chosen to reflect popular avenues of explainability within human-robot or human-AV interactions~\cite{Paleja2021Minecraft,huber2021local,ehsan2019automated,omeiza2022}.

\subsection{Procedure}
\label{subsec:phase1-procedure}
Upon arrival to the onsite location, participants complete consent forms and are briefed on their task. Participants are introduced to each of the xAI mechanisms employed in the study, the interface for directing the car, and a mini-map that will assist them for each task. They then complete the Negative Attitudes towards Robots Scale (NARS) \cite{nars}, ``Big-Five'' personality \cite{mini-ipip}, and demographic data surveys, used as controls in our statistical analyses. Participants then begin on eleven navigation tasks (Section \ref{subsec:domain}). 

Participants complete two practice tasks to become acquainted with the simulator, controls, and explanations. Pilot studies revealed that very little practice was required for the task, so two tasks was sufficient. In this practice phase, explanations are randomly sampled from any of the three mechanisms used in our study (Section \ref{subsec:driving-agent}), giving the participant equal practice with each modality.

After completing the practice phase, participants begin the main body of the study, which consists of nine navigation tasks. Each task uses a single xAI modality from start to finish, which helps to mitigate consecutive mistakes that may stem from swapping between xAI modalities. The agent rotates between modalities as tasks are completed, and the ordering of xAI modalities is included as a control in our statistical analyses.
Participants conclude the study with a survey asking them to rank the three xAI modalities according to their preferences. 

\subsection{Results}
\label{subsec:phase1-results}
The population study involved 30 participants (Mean age = 23.8, SD = 3.25; 70\% Male).
\begin{figure*}[!ht]
  \centering
  \includegraphics[width=\textwidth]{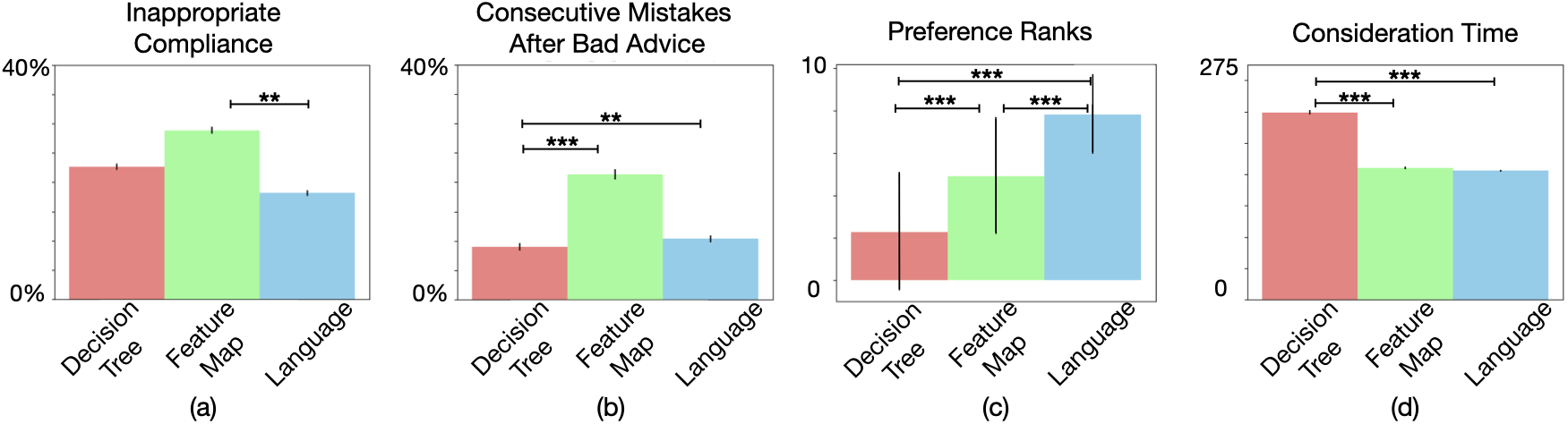}
    
  \caption{Visualized results from the population user study between decision trees, feature-importance maps, and language explanations. (a) Feature maps lead to significantly increased inappropriate compliance. (b) Both feature maps and language explanations lead to more consecutive mistakes (quantities are normalized by total number of mistakes). (c) Language is significantly preferred over decision trees and feature maps. (d) Decision trees are slower to parse (measured in seconds).}
  \label{fig:phase1-results}
  \vspace{-0.25cm}
\end{figure*}

\textbf{RQ 1.1 --} Comparing the sum across five Likert-items as the preference rank, an ANOVA for xAI modality rankings showed a significant difference across baselines ($F(2, 84) = 35.1, p < 0.001$). A Tukey-HSD revealed that language explanations ranked significantly higher than both feature-importance maps \statcheck{($p < 0.001$)} and decision trees \statcheck{($p < 0.001$)}, and feature-importance maps ranked significantly higher than decision trees \statcheck{($p < 0.001$)} (Figure \ref{fig:phase1-results}). 

\textbf{RQ 1.2 --} Data for inappropriate compliance did not pass a Shapiro-Wilk test for normality, and we therefore applied a Friedman's test, which was significant \statcheck{($\chi^2(2)=12.23, p = 0.002$)}, with a post-hoc revealing that language explanations lead to significantly fewer instances of inappropriate compliance than feature-importance maps \statcheck{($p = 0.002$)} (Figure \ref{fig:phase1-results}). 

Data for consideration time were not normally distributed. We therefore applied a Friedman's test, which was significant \statcheck{($\chi^2(2)=24.47, p < 0.001$)}. A post-hoc revealed that both language and feature-importance explanations are significantly faster than decision-tree explanations \statcheck{($p < 0.001$)}.

Finally, an ANOVA across explanation modalities for consecutive mistakes was significant \statcheck{($F(2, 74) = 8.0309, p < 0.001$)}, and a post-hoc revealed that participants make significantly more consecutive mistakes with feature-importance maps \statcheck{($p < 0.001$)} and language explanations \statcheck{($p = 0.001$)} than with decision trees. 

\textbf{RQ 1.3 --} After grouping participants by their preferred modality, we do not find any significantly different trends for performance (i.e., participants that favor feature-importance maps do not perform best with feature-importance maps).

\subsubsection{Takeaways}
\label{subsubsec:phase1-takeaways}
A review of the results for the population study reveals that language explanations are both significantly preferred relative to feature-importance and decision-tree explanations, and result in higher task-performance than feature-importance explanations. We do find, in line with contemporary work~\cite{tambwekar2023towards}, that decision trees result in significantly fewer consecutive mistakes, suggesting that it is easier for participants to reflect on why the previous decision was incorrect and to immediately update their mental model of the agent. However, across most metrics, language explanations are superior to both other modalities considered in this work. We therefore consider language explanations to be the ``gold standard'' for this task (i.e., an agent that presents only language explanations will be significantly preferred and yield significantly higher task-performance for most of the population).

\section{Personalization Study}
\label{sec:phase2}

The population study helped identify a ``gold-standard'' explanation for our domain, and revealed significant population-wide trends with regards to preference and performance. 
However, knowledge of the best xAI modality is not readily available for most domains, 
and there may be adverse effects of universally applying population-wide trends on an individual level~\cite{personal_driving_state}.
Our personalization study therefore studies the effects of different personalization strategies on preference- and task-performance-maximization, including a novel adaptive personalization approach that balances between a participant's preferences and task-performance needs\footnote{The xAI modalities are adjusted slightly between the population and personalization studies, following feedback from some population-study participants that the explanations were too simplistic and easy to memorize. Each modality was therefore made slightly more complicated, and an additional 12 pilot participants verified that the new explanations fairly reflected the results of the population study, while increasing in complexity. Details and examples are available in the supplementary material.}.

\subsection{Adaptive Personalization Approach}
\label{subsec:phase2-technical}
The population study revealed that preference and performance do not necessarily align. This observation inspired the development of an adaptive personalization technique that can balance between a participant's preferences or performance-needs depending on the situation. To accomplish this personalization, our agent must model preferences or performance for each participant.

As discussed in Section \ref{subsec:shared-metrics}, the agent tracks how often participants make mistakes (i.e., do not follow the optimal path), as well as the participant's feedback, for each xAI modality. Using these two metrics, the agent creates a preference distribution and a task-performance distribution for the participant. These distributions are largely driven by \textit{negative} interactions with the agent (i.e., negative feedback or mistaken turns). The decision to focus on negative feedback owes to pilot studies, which revealed that negative interactions are rarer and more meaningful than positive interactions.

\subsubsection{Producing Sampling Distributions}
\label{subsubsec:phase2-distributions}
First, the agent counts all interactions for each modality and stores the resulting values in a vector, $\vec{x}$ (e.g., counting the total number of language, feature-map, and decision-tree interactions). The agent then separates this out into two additional quantities-- the total number of negative interactions, $\vec{x_{-}}$ and the total number of positive interactions $\vec{x_{+}}$. The basis of the sampling distribution is then computed as $\vec{x_{-}} * \frac{\vec{x}}{\vec{x_{+}}}$. In other words, the total number of negative interactions for each modality, smoothed by the ratio of total-to-positive interactions. This normalized quantity will be high for modalities where interactions are more often negative, and low for xAI modalities that have far more positive than negative interactions. 

The agent tallies the number of modalities with at least one negative interaction, which normalizes the above quantity, smoothing the distribution if negative interactions occur in all modalities. Finally, this value is negated so that modalities with higher negative values will be sampled less frequently. 
The distribution over all xAI modalities is computed according to Equation \ref{eqn:dist-neg}.
\begin{equation}
    \label{eqn:dist-neg}
\vec{v} = -\frac{\vec{x}_{-}*\frac{\vec{x}}{\vec{x}_{+}}}{\sum_{i=0}^{|\vec{x}|} (1 \, \text{if} \, \vec{x}_{-, i}>0)}
\end{equation}

The resulting distribution, $\vec{v}$, is then normalized using a softmax function to produce a probability distribution. When $\vec{x}$ is the vector of task-performance interactions (i.e., correct and incorrect turns), we obtain a distribution for task-performance, $\vec{d}_T$. If $\vec{x}$ is instead a vector of feedback interactions, we obtain a distribution of the participant's preferences, $\vec{d}_P$. Sampling from $\vec{d}_P$ will maximize the likelihood of selecting an explanation that aligns with satisfying a participant's preferences, while sampling from $\vec{d}_T$ will maximize the likelihood of selecting an explanation that helps the participant to identify the optimal action. However, there is no notion of balancing between these two, potentially-competing, objectives.

\subsubsection{Balancing Between Multiple Objectives}
\label{subsubsec:phase2-balancing}
To achieve the balance between participant preference and task-performance, we must find a way to balance between $\vec{d}_P$ and $\vec{d}_T$ depending on the participant's progress in the task. To this end, we define a new distribution, $\vec{d}_B$, that balances between $\vec{d}_P$ and $\vec{d}_T$, using a trade-off parameter $\lambda$. The trade-off parameter should emphasize $\vec{d}_P$ if the participant is going to be correct (i.e., adhere to participant preferences if there is little risk of a mistake), and emphasize $\vec{d}_T$ if there is a high risk of the participant being incorrect (i.e., ignore preferences and maximize task-performance if a mistake is likely). $\vec{d}_B$ is therefore constructed according to Equation \ref{eqn:sample}.
\begin{equation}
    \label{eqn:sample}
    \vec{d}_{B} = \lambda * \vec{d}_{P} + (1-\lambda) * \vec{d}_{T}
\end{equation}

To obtain this trade-off parameter, $\lambda$, the agent requires an estimate of whether the participant is likely to make a mistake. To this end, the agent employs a neural network to predict which direction the participant will choose at each intersection. This network consumes state information from the environment (e.g., position, orientation, goal position, and nearest roadblock position), and is trained over data collected from pilot studies and the population study (Section \ref{sec:phase1}). However, because participants often take different paths, this network must personalize to each participant. The network therefore maintains a unique embedding for each participant, following prior personalization research \cite{schrum2022mind,silva2022fedembed,paleja2020interpretable}. Similarly, the network maintains a unique embedding for each navigation task, allowing for contextual adaptation in addition to individualized personalization, as in \cite{fedperc}. These embeddings are both passed into the network alongside state information. During deployment the main body of the network is frozen, and only the personal and contextual embeddings are updated as participants act in the domain.

At each intersection, this model predicts which direction the participant is going to choose, producing a probability distribution over the three possible directions, $\vec{y}$. If the agent predicts that the participant is going to go in the optimal direction, then the trade-off parameter, $\lambda$, is set to $argmax(\vec{y})$ (i.e., set to the logit value for the optimal direction). Otherwise, $\lambda$ is set to $1-argmax(\vec{y})$.

\subsection{Study Conditions}
\label{subsec:phase2-conditions}

The personalization study compares five xAI-selection strategies:
 \begin{itemize}
     \item \textbf{Balanced personalization} -- explanations are drawn from $\vec{d}_B$, as described in Section \ref{subsec:phase2-technical}.
     \item \textbf{Preference maximization} -- explanations are drawn from $\vec{d}_P$, only conditioning on participant preferences.
     \item \textbf{Task-performance maximization} -- explanations are drawn from $\vec{d}_T$, only conditioning on participant task-performance.
     \item \textbf{Random explanations} -- explanations are randomly selected from the three available modalities.
     \item \textbf{Language-only explanations} -- all explanations use the language condition, known a-priori to yield the best performance and match most people's preferences (Section \ref{subsubsec:phase1-takeaways}).
 \end{itemize}

\subsection{Procedure}
\label{sec:pers_procedure}
Following a pilot study, the personalization study is designed as a set of within-subjects experiments. 
In this study, participants work with two personalization strategies (Section \ref{subsec:phase2-conditions}).
 Each experiment compares balanced-personalization to one other approach for choosing explanations.
 This design helps to control for variance across participants, which we observed to be high.
 The ordering of these two strategies is counter-balanced across all participants.
 
 The personalization study begins by following the same procedures as in the population study (Section \ref{subsec:phase1-procedure}). After being briefed on the task and the xAI modalities, participants complete  a single training task and then begin two calibration tasks. All three tasks (one training and two calibration) randomly cycle through all available xAI modalities, giving participants exposure to the various explanations they will receive. Over the two calibration tasks, the agent begins to gather feedback and observations on the participants behavior to create $\vec{d}_P$ and $\vec{d}_T$. The agent also uses these tasks to learn personal and contextual embeddings for the personalization network (Section \ref{subsubsec:phase2-balancing}).

 After calibration tasks, the participants complete three navigation tasks with the first selection strategy, and then stop to complete a set of surveys on trust \cite{jian2000foundations}, perceptions of social competence \cite{bartneck2009measurement}, perceived workload \cite{TLX}, and explainability \cite{silva2022explainable}. Participants then provide free-form text on how well they thought that the agent conformed to their preferences. After this set of questions, participants resume the navigation tasks, completing an additional three tasks with the second selection strategy. Finally, participants complete the same set of questionnaires a second time.

\begin{figure}[t]
  \centering
  \includegraphics[width=\linewidth]{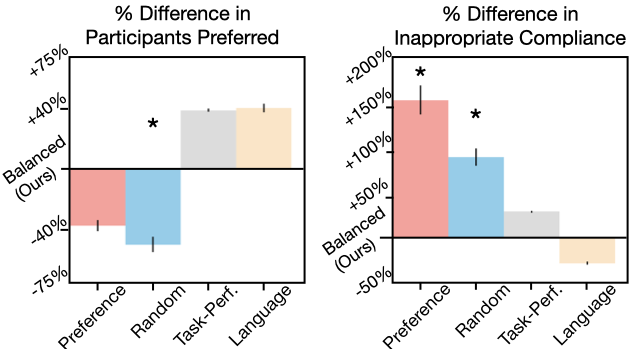}

  \caption{Comparisons relative to balanced-personalization. (Left) Percent of participant preferences for personalization modes, showing significant preference for balanced personalization over no personalization. (Right) Rates of inappropriate compliance, showing that balanced-personalization leads to significantly lower inappropriate compliance than preference-maximization or no personalization.}
  \label{fig:phase2-results}
    \vspace{-0.5cm}
\end{figure}

\subsection{Results}
\label{subsec:phase2-results}
 The personalization study involved 60 participants (Mean age=24.87, SD=6.75; 53\% Male). While a subset of significant results are presented here, all statistical test details and pairwise condition comparisons are presented in full supplementary material.

\textbf{RQ 2.1 --}
Comparing balanced personalization to random explanations, a wilcoxon signed rank test for preference rankings showed a significant difference across conditions \statcheck{($W=4, p=0.04$)}. Additionally, comparing binary feedback ratings for preference maximization and balanced-personalization, a Friedman's test revealed significantly higher feedback ratings for preference-maximization \statcheck{($\chi^2(1)=5.444, p=0.02$)}. Interestingly, we find that, while preference-maximization leads to significantly higher positive binary feedback during the study, it appears to result in lower retrospective preference ratings after the study (Figure \ref{fig:phase2-results}). 


We do not find statistically significant differences between either feedback data or text-responses for the task-performance agent vs. the balanced-personalization agent. We therefore find evidence to answer \textbf{RQ 2.1}-- balanced personalization is significantly preferred over random explanations.

\textbf{RQ 2.2 --} Comparing balanced-personalization to random explanations, a wilcoxon signed rank test reveals significantly higher inappropriate compliance using random explanations \statcheck{($W = 25, p = 0.03$)}. We also observed significantly higher inappropriate compliance in the preference-maximization agent compared to balanced-personalization \statcheck{($W = 21, p = 0.02$)} (Figure \ref{fig:phase2-results}). Similarly, participants took significantly more steps above optimal performance with a preference-maximization agent compared to a balanced-personalization agent  \statcheck{($W=31, p=0.04$)}. We therefore answer \textbf{RQ 2.2}-- balanced personalization yields significantly higher performance relative to preference-maximization or random explanations.


\textbf{RQ 2.3 --} We find no statistically significant differences between the balanced-personalization agent and language-only agent along the performance or preference metrics.  
We therefore find that balanced personalization is not worse along task-performance and preference metrics when compared to the a-priori known best xAI modality for our domain.

\subsubsection{Takeaways}
\label{subsubsec:phase2-takeaways}
Reviewing the results of the personalization study, we find that balanced personalization is significantly superior to no personalization (i.e., random explanations) along the axes of both preference and task-performance. Similarly, balanced personalization leads to significantly fewer mistakes than preference maximization. We find no significant differences between task-performance maximization, suggesting that task-performance is of paramount importance for participants in our study \cite{yang2017evaluating}. In other words, our study found that receiving explanations the participants did not prefer to use (e.g., seeing mostly decision trees even when asking to stop receiving them) did not register as the agent not conforming to participant preferences, so long as the participant was perceived to be making progress on the task. This trend could potentially be due to ``experienced accuracy'', wherein a participant's experience or perception of an agent's accuracy affects their interactions with the system~\cite{yin2019understanding, liao2020questioning}. 

Finally, we find no statistically significant differences between language-only explanations (known to be best before the study) and a balanced-personalization agent. This finding implies that balanced personalization will not under-perform the best xAI mode for a new domain, while avoiding the need to run a population study. Deploying balanced personalization can significantly reduce the overhead for deploying xAI to new domains while also ensuring that participants receive xAI modalities that match their needs (e.g., feature-importance maps for participants that cannot use language).

\section{Discussion and limitations}
In the population study, we find significant differences between the three xAI modalities, decision trees, language, and feature-importance maps, examined in this work when considering participant preference and task-performance. Despite these trends, we find an interesting counterexample, in which decision-tree explanations are significantly better for identifying mistaken decision-making processes for consecutive errors (Section \ref{subsec:phase1-results}). This result echoes prior work \cite{tambwekar2023towards}, finding that language explanations were significantly preferred by untrained participants, but that participants were better at modeling agent behavior when using decision trees.

In the personalization study, we confirm that balanced personalization yields significant performance improvements relative to preference maximization or no personalization. Furthermore, while participants provide significantly more negative responses to a balanced-personalization agent, they retrospectively perceived the balanced-personalization to do a \textit{better} job conforming to their preferences. Similarly, we find that a task-performance maximization agent receives very positive retrospective perceptions of personalization (40\% over balanced-personalization, shown in Figure \ref{fig:phase2-results}), despite \textit{never} considering $\vec{d}_P$ when making decisions. Together, these findings suggest that participants in the personalization study prized task performance over preference-accommodation. Until a satisfactory level of performance is met, accommodating preferences may not be perceived as important or useful, as participants seem to fixate on optimally completing the task rather than on engaging with an agent that listens to their feedback. This finding echoes prior work on the effects of performance on trust \cite{yang2017evaluating}, and underscores the importance of personalizing xAI for task performance. Applying a balanced personalization approach, as introduced in this work, we can achieve the benefits of maximizing task-performance at crucial junctions (e.g., if the user is likely to make a mistake or if their mental model appears to be incorrect) while \textit{also} accommodating user preferences.

While our work was conducted on a simulated task, our findings generalize more broadly to any domain in which a human might need to work with concept-based explanations \cite{das2023subgoals}, such as feature-maps, decision trees, or counterfactual explanations. Using such explanations, a mistake is often identified by the inclusion of an errant feature, as in this research. We show that humans have diverse preferences and experiences when working with such explanations, and that adaptive personalization can enhance human-robot interactions that rely on xAI for decision-verification \cite{hayesshah2017}.

\textit{Limitations -- } These studies were conducted primarily with university students on a driving simulator, rather than on an autonomous vehicle with a broader population. Additionally, we observe some results with large effect sizes but no statistical significance (Figure \ref{fig:phase2-results}), which may be due the sample size in our studies. 

Personalization agents in this work also assumed access to ground truth information in the domain (e.g., optimal turns) or explicit preference feedback, which may be challenging to obtain the real-world. While such ground-truth directions could come from external sources (GPS navigation) and feedback data could come from observations of humans \cite{axelsson2023}, these external data sources are not used in our work. Finally, personalization in this work was confined to selecting xAI modalities that match a participant's preferences, but did not extend to adapting the explanations themselves. 

\section{Conclusion}
To be useful in the real world, digital agents and robots must be able to personalize to a diverse population of users, even without prior knowledge of the best way to interact or the most popular interaction modality for a given task. In this work, we have studied the differences between three popular explainability techniques: language explanations, feature-importance maps, and decision trees, in the context of a simulated AV study. We have presented an approach to personalization that balances subjective human-preference with objective task-performance. A separate user study confirmed that such a balanced personalization approach yields significantly improved task-performance relative to a preference maximization agent, and is not worse than an agent that uses explanations which are known a-priori to maximize preference scores and task-performance in the population. Our study is the first to personalize explanations to task-performance, and we show that personalization must consider task-performance to be successful. We discuss the implications of our results, including the need for high-performance agents before considering preference maximization, and the need to carefully balance adherence to preferences with task-performance. 

\section*{Acknowledgements}
This work was supported by NSF award CNS 2219755, NASA Early Career Fellowship 80NSSC20K0069, and a gift from Konica Minolta.

\bibliographystyle{ACM-Reference-Format}
\balance
\bibliography{sample-base}

\begin{figure*}
  \includegraphics[width=\textwidth]{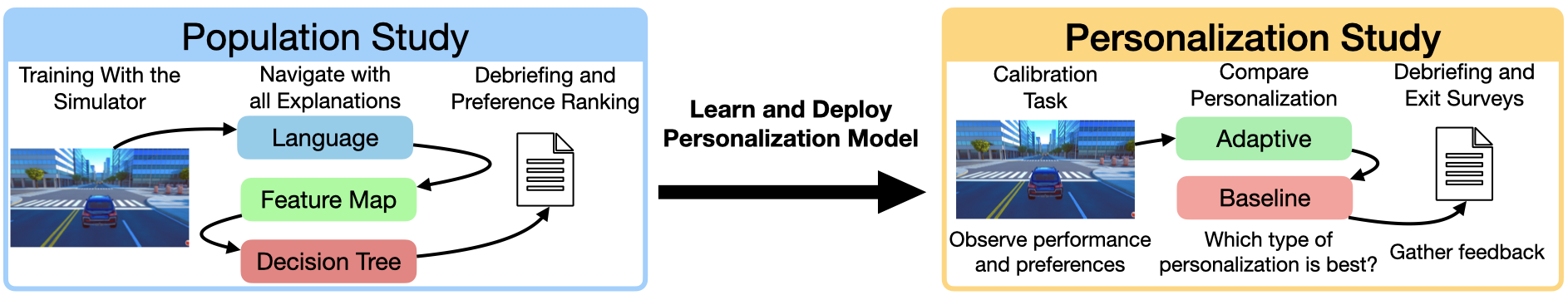}
  \caption{We conduct two user studies, beginning with a population study (left) in which all participants work with three xAI modalities, revealing significant differences across the population. We then use this data to build an adaptive personalization model that is deployed in a set of personalization studies to compare various personalization approaches.}
  \label{fig:pipeline}
\end{figure*}

\begin{figure*}[t]
  \centering
  \subfloat[][]{\includegraphics[width=.2\textwidth]{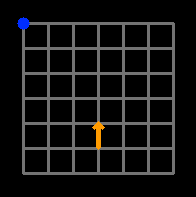}} \quad
  \subfloat[][]{\includegraphics[width=.5\textwidth]{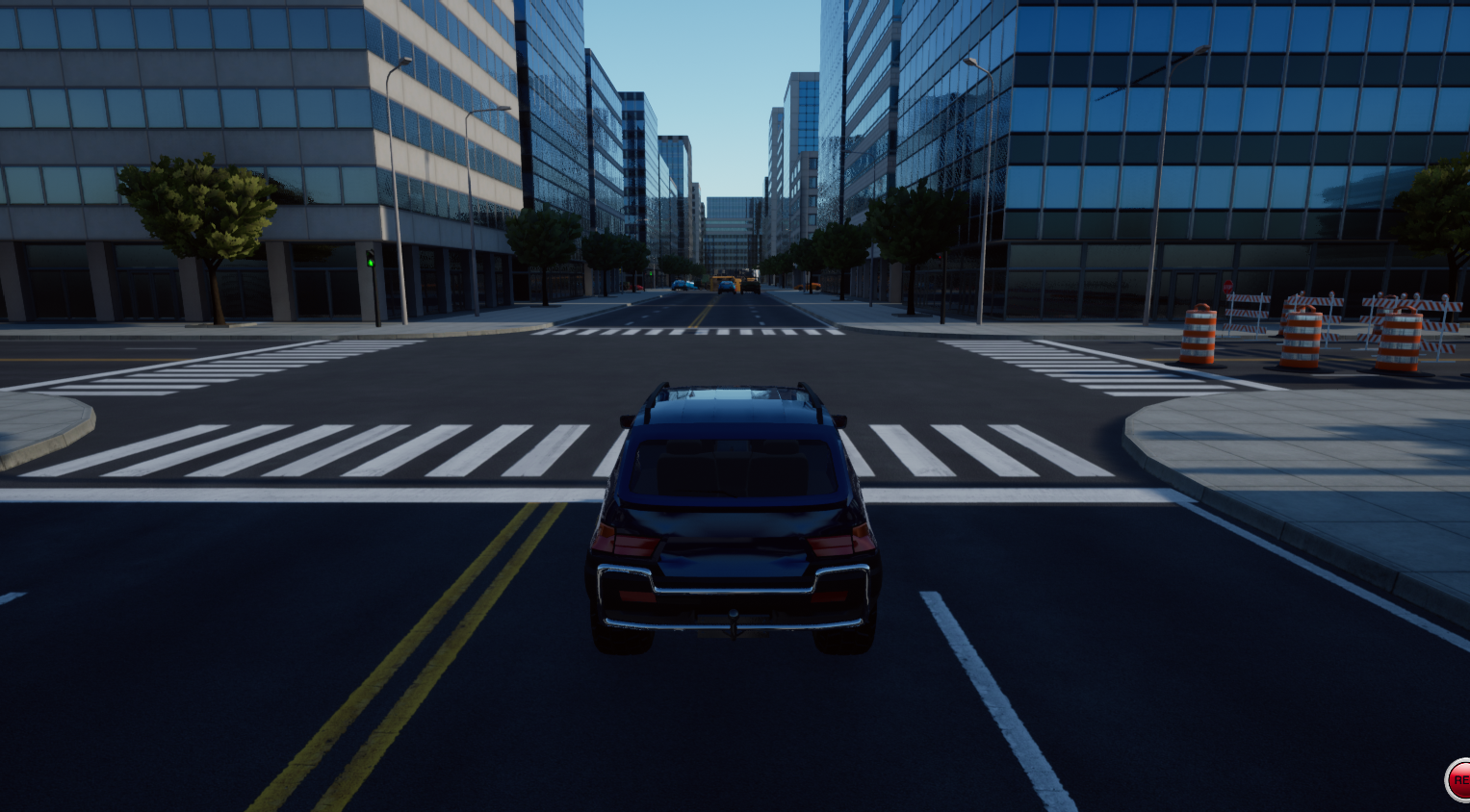}} \\
    \subfloat[][]{\includegraphics[width=.7\textwidth]{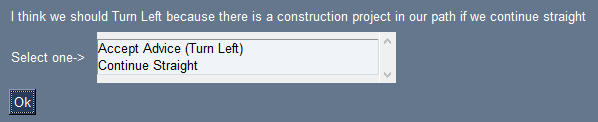}} \\

  \caption{Our custom self-driving domain created using Unreal and the AirSim \cite{shah2018airsim} simulator. At each intersection, each participant is shown a mini-map of the city (a), in order to assist them in their decision making. The mini-map provides the location and heading of the car, as well as the location of the goal. Participants select a direction from a pop-up to direct the car. In this example, the pop-up includes a language explanation.}
  \label{fig:domain}
\end{figure*}

 \section{Additional Domain Details}
 \label{app:domain-details}
 Before and after each study, participants complete several surveys including: demographics, negative attitudes towards robots \cite{nars}, mini-IPIP personality survey \cite{mini-ipip}, and experience with driving, robots, and decision trees. Following the completion of a portion of the studies in this work, participants complete a robot trust survey \cite{jian2000foundations}, the NASA-TLX survey of perceived workload \cite{TLX}, a survey on perceptions of anthropomorphism and social competence of the digital agent \cite{bartneck2009measurement}, and a survey on the perceived explainability of a digital agent \cite{silva2022explainable}.

 \section{Overall Study Flow}
 We provide an overview figure for our study flow in Figure \ref{fig:pipeline}. In our population study, participants first complete consent forms and are briefed on the task. They then fill out pre-surveys and demographic information, before training on the simulator for two tasks. After the training phase, participants being the main body of the study, rotating through each of the available conditions for a total of 3 tasks with each xAI modality (e.g., language, then feature-importance, then decision-trees, then repeat the cycle two more times). After their final task, participants provide their preference rankings for each of the xAI modalities, and are finally debriefed on their experience.

 In the personalization study, participants also begin with consent forms, briefing, pre-survey, and demographic surveys. Participants then complete one training task and two calibration tasks, in which the driving agent rotates through each xAI modality for every intersection. After the second (and final) calibration task, participants are randomly assigned to either the adaptive personalization strategy or to a baseline condition. Participants complete three tasks with this strategy, then fill out a preference survey. After the preference survey, participants resume the study with the other personalization strategy (either adaptive or baseline). After completing three more tasks, participants redo the preference survey for the second personalization strategy, and are then debriefed.

\section{Providing Incorrect Suggestions and Explanations}
\label{app:incorrect-explanations}
In both studies, incorrect suggestions were provided as the exact \textit{opposite} of the correct direction, and participants were warned of this information at the beginning of the study. We opted to make incorrect suggestions point in the opposite direction (as opposed to a random incorrect direction) so that participants could, in theory, always take the optimal route (e.g., if the agent is incorrect and says ``go left'', the participant knows that going ``right'' is optimal). 

Because there are often three available directions, choosing an ``opposite'' is possible for only two out of three options (i.e., left and right). When the correct direction is simply ``straight'', we reduce the number of options by arbitrarily disabling one direction. In other words, the agent randomly masks out ``right'' or ``left'', thereby only presenting two options to the participant. The incorrect suggestion is whichever direction was not masked (``left'' or ``right''), and the correct direction is to go in the only other option available (i.e., ``straight''). Participants are told how to handle this situation during the briefing (Appendix \ref{app:briefing}).

In practice, we found that many participants did pick up on these rules, and did not struggle to know how to handle an explanation that they perceived to be incorrect (though they did not always understand when or why explanations were incorrect). Commonly, participants struggled when they knew a suggestion was incorrect, but they wanted to accept the suggestion because the direction itself seemed to be correct from their viewpoint. For example, consider the situation where the goal is immediately to the participant's left, but a construction site lay between the participant and the goal. A participant that wants to get to the goal as quickly as possible will want to turn left. If the digital assistant incorrectly suggests ``left'' and provides an invalid explanation, the participant may recognize that they \textit{should not} go left, but they may turn left anyway, simply because they already expected they should go that way (and often, they hoped that ``this time it will be right,''). Similarly, if the agent \textit{correctly} suggested an alternate direction, participants may recognize that they \textit{should} comply (i.e., take a less direct path, such as going ``right'' in the above example), but still end up going the wrong way, simply because they hope that their more direct path will work. 

\begin{figure*}
  \includegraphics[width=\textwidth]{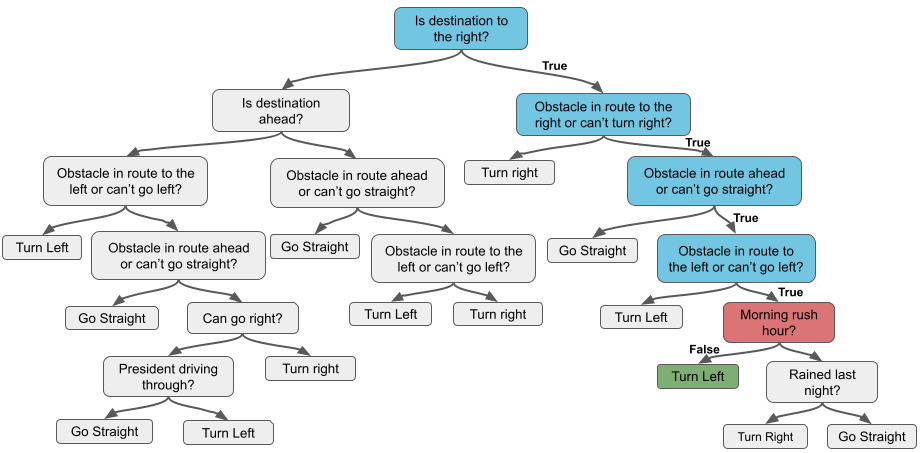}
  \caption{A decision tree explanation that the participant should ignore. Note that the highlighted path (i.e., the decision suggestion) includes a red-herring feature (rush-hour traffic).}
  \label{fig:incorrect_dt}
\end{figure*}

\section{Incorrect Explanations}
\label{app:incorrect}
Incorrect explanations were signalled by the inclusion of ``red-herring'' features, as told to participants (Section \ref{app:briefing}). These included: the weather, the radio, the sky, traffic, rush hour, or the president's motorcade. Participants are explicitly told most of these (they are not explicitly told about the president's motorcade, though they do see it in the training and calibration tasks, so they are able to learn that rule before beginning the main task). They are also explicitly told that any explanation considering ``external factors'' (i.e., not pertaining to the road or to construction sites and car crashes) is an incorrect explanation. An example of an incorrect decision tree explanation is given in Figure \ref{fig:incorrect_dt}.

Examples of incorrect explanations in our work included:
\begin{itemize}
    \item ``I think we should turn left because the president's motorcade is in town.''
    \item ``I think we should continue straight because it didn't rain last night.''
    \item ``I think we should turn right because clear skies could impair our cameras.''
\end{itemize}

\begin{figure*}
  \includegraphics[width=\textwidth]{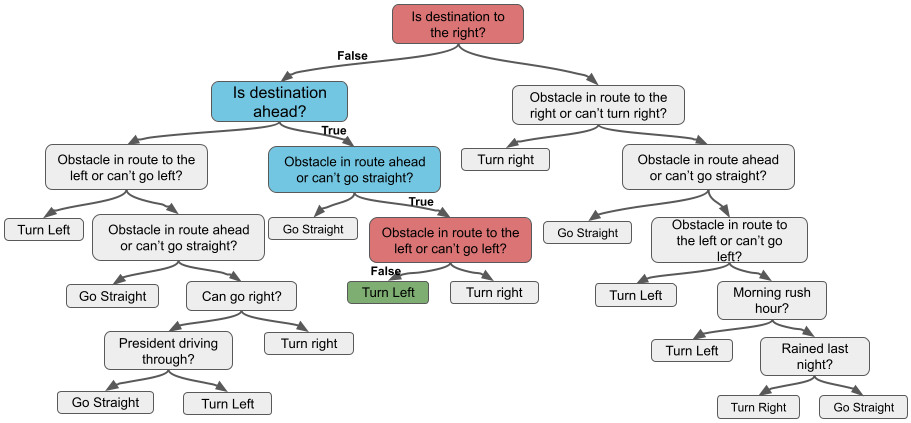}
  \caption{A decision tree explanation that the participant should adhere to. Note that the highlighted path (i.e., the decision suggestion) considers only relevant details (path to the goal and obstacles) and ignores red-herring features.}
  \label{fig:correct_dt}
\end{figure*}

\section{Correct Explanations}
\label{app:correct}
Correct explanations were signalled by failing to include ``red-herring'' features. In particular, if the explanation pertained only to the ``shortest path'' or ``optimal route'', then it was correct. Additionally, if the explanation contained only information about the goal or obstacles on the path to the goal (e.g., construction sites or car crashes), then it was correct. For feature-importance maps, which did not necessarily contain any of this information, explanations were correct as long as they did not highlight the sky. An example of a correct decision tree explanation is given in Figure \ref{fig:correct_dt}.

Examples of correct explanations in our work included:
\begin{itemize}
    \item ``I think we should turn left because there is a construction project in our path if we turn right.''
    \item ``I think we should continue straight because it is the shortest path to the goal.''
    \item ``I think we should turn right because we will hit a pile-up if we continue straight.''
\end{itemize}

\section{Explanation Changes Between Studies}
\label{app:between-study-changes}
After the conclusion of the population study, we opted to change the content of some of the explanation modalities. This is because a few participants mentioned that they found it easier to memorize all \textit{correct} explanations, rather than learning the rule to identify \textit{incorrect} explanations (particularly for the language modality). However, the goal of our work is to study when xAI enables users to identify errant decision-making from a digital assistant, not to study how easy it is to memorize all possible correct explanations (which would be impractical in a real-world setting).

\begin{figure*}[!ht]
  \centering
  \includegraphics[width=0.7\textwidth]{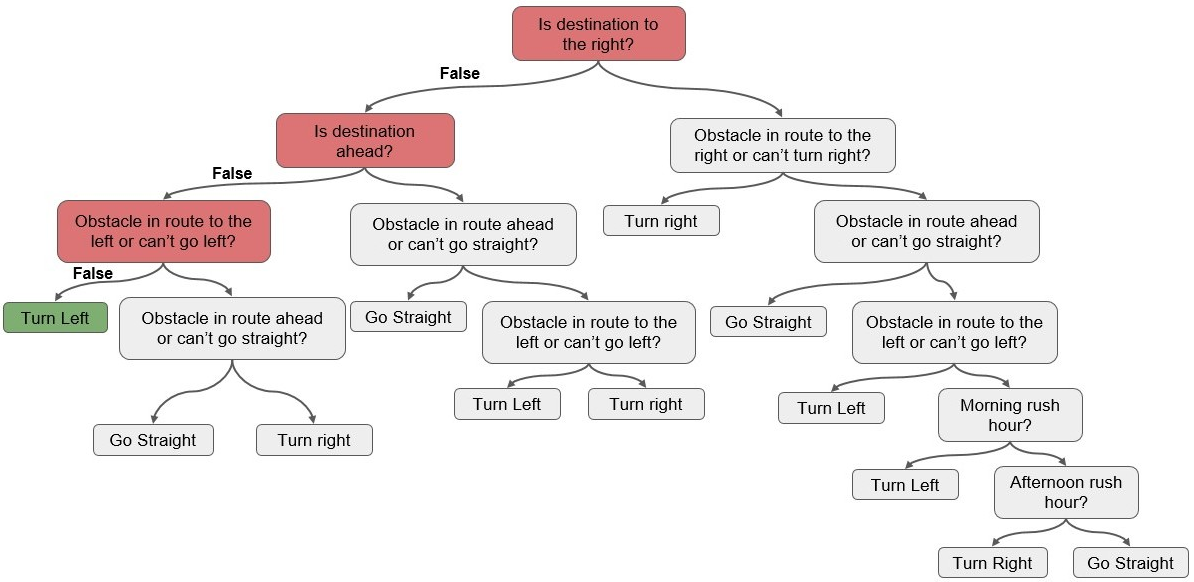}

  \caption{Decision tree explanation from the population study.}
  \label{fig:old_dt}
\end{figure*}

\begin{figure*}[!ht]
  \centering
  \includegraphics[width=0.7\textwidth]{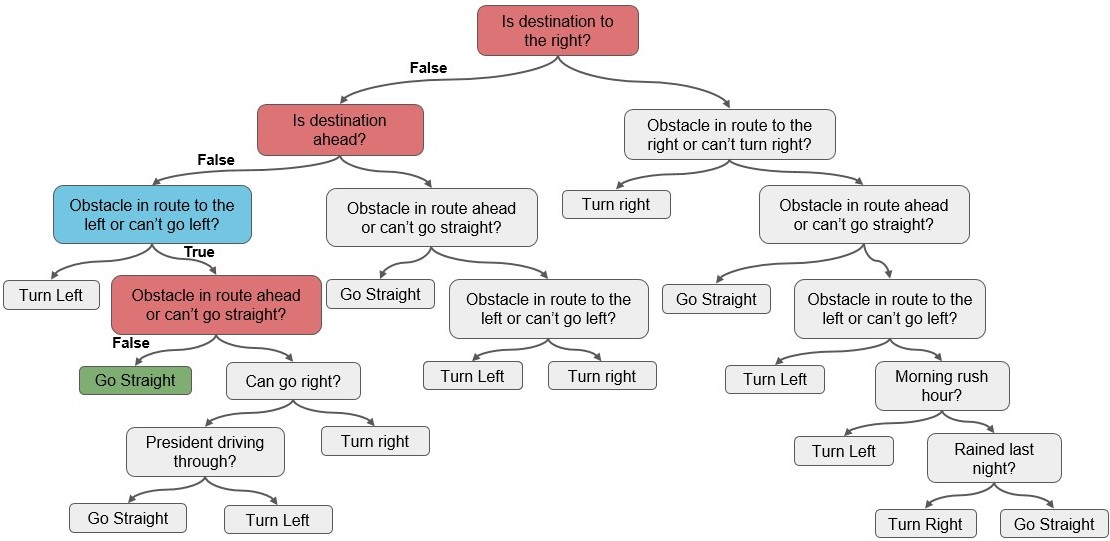}

  \caption{Decision tree explanation from the personalization study, with one decision node and 2 leaf nodes added.}
  \label{fig:new_dt}
\end{figure*}

To make the explanations more complex, we added several language templates and rephrases, thereby greatly increasing the number of possible language explanations (from 6 up to 47). We also added one new decision node and 2 new leaf nodes to the decision tree. For reference, the original decision tree is presented in Figure \ref{fig:old_dt}, and the updated tree is presented in Figure \ref{fig:new_dt}. Finally, for feature-importance explanations, we modulated the brightness (i.e. importance) of buildings and trees in the image. Rather than being set to a static color, the color and brightness was changed to be randomly sampled, with a low set of values defined for correct explanations (Figure \ref{fig:right_feature_map}), and a high set of values defined for incorrect explanations (Figure \ref{fig:wrong_feature_map}). We conducted a pilot study with 12 additional participants using these new explanations, which showed that there were no significantly different trends from the population study. Upon examining the results and debriefing participants, we found that no participant was able to memorize correct explanations after these changes.

\section{Task Orderings}
\label{app:orderings}
In the population study, participants were required to complete nine navigation tasks, three with each explanation modality. Participants rotated between explanation modalities, such that every third task was completed with the same explanation modality. We enumerated all six possible orderings of explanation modes (e.g., (1) decision-tree, (2) feature-importance, (3) language, or (1) feature-importance, (2) language, (3) decision tree, etc.) and distributed participants evenly across all orderings, such that each ordering received five participants.

In the personalization study, we included a total of six ``test'' tasks (i.e., not training or calibration). All participants went through the same six test tasks. We constructed a Latin square to order tasks for participants, running the studies with balanced orderings. We also balance the ordering of which explanation selection strategy is shown first or second, resulting in a study with 12 participants for each comparison. An additional 12 participants (for a total of 24) are recruited for the task-performance maximization vs. balanced personalization study (i.e., $\vec{d}_T$ vs. $\vec{d}_B$), following the results of a power-analysis.

\subsection{Statistical Test Details}
\label{app:stats-test-details}
We performed a repeated-measures multivariate analysis to compute the effects of different conditions (explanation modality in the population study, personalization approach in the personalization study) on various metrics (explanation preference ranking, inappropriate compliance, etc.). Across various tests, the condition is modeled as a fixed-effect covariate, participant ID is a random effect covariate. We use the AIC metric to determine which additional covariates should be included for each test, considering task ordering, condition ordering, and different resposnes from a demographic data pre-survey (e.g., race, gender, age, robotics experience, etc.). We then apply an analysis of variance (ANOVA) to identify significance across baselines, and further employ a Tukey-HSD post-hoc test to measure pairwise significance. For our linear regression model, we tested for the normality of residuals and homoscedasticity assumptions. If the data do not pass normality assumptions, we apply a non-parametric Friedman's test with a Nemenyi's All-Pairs Comparisons post-hoc. If the data do not pass homoscedasticity assumptions, we proceed with a wilcoxon signed rank test. Finally for binary data and count data, we apply a wilcoxon signed rank test. 

\section{Additional Results}
\label{app:additional-results}
We present full pairwise comparisons between conditions in this section, showing the results when controlled for variance across participants in different conditions.
\begin{figure*}[!ht]
  \centering
  \includegraphics[width=\textwidth]{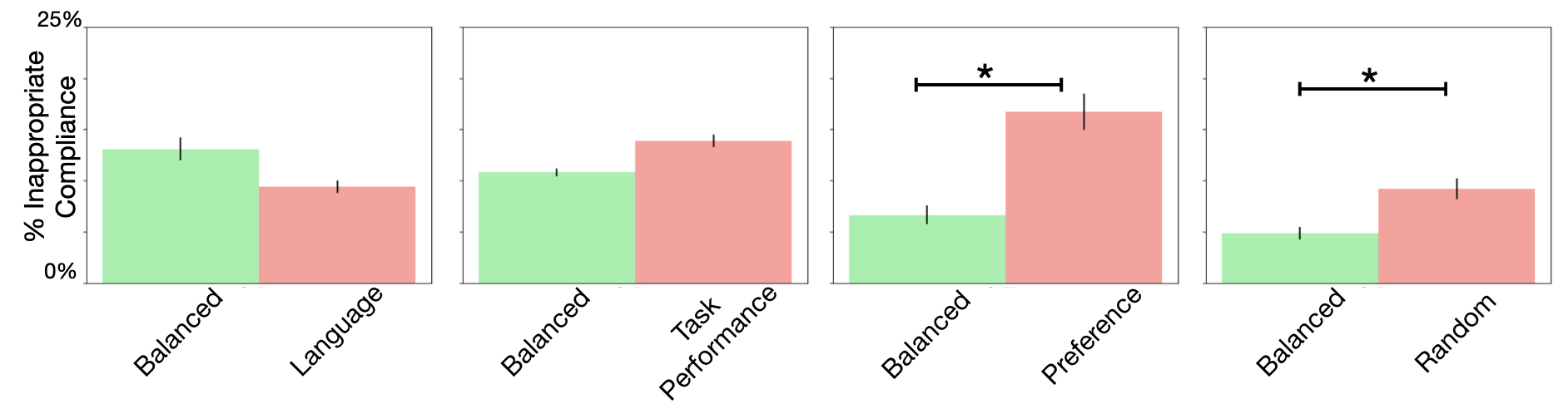}

  \caption{Visualized percent of inappropriate compliance across all condition-comparisons in the personalization study. Balanced personalization helps participants identify errant decision suggestions significant more than preference maximization or no personalization at all.}
  \label{fig:phase2-wrong_compliance}
\end{figure*}
\begin{figure*}[!ht]
  \centering
  \includegraphics[width=\textwidth]{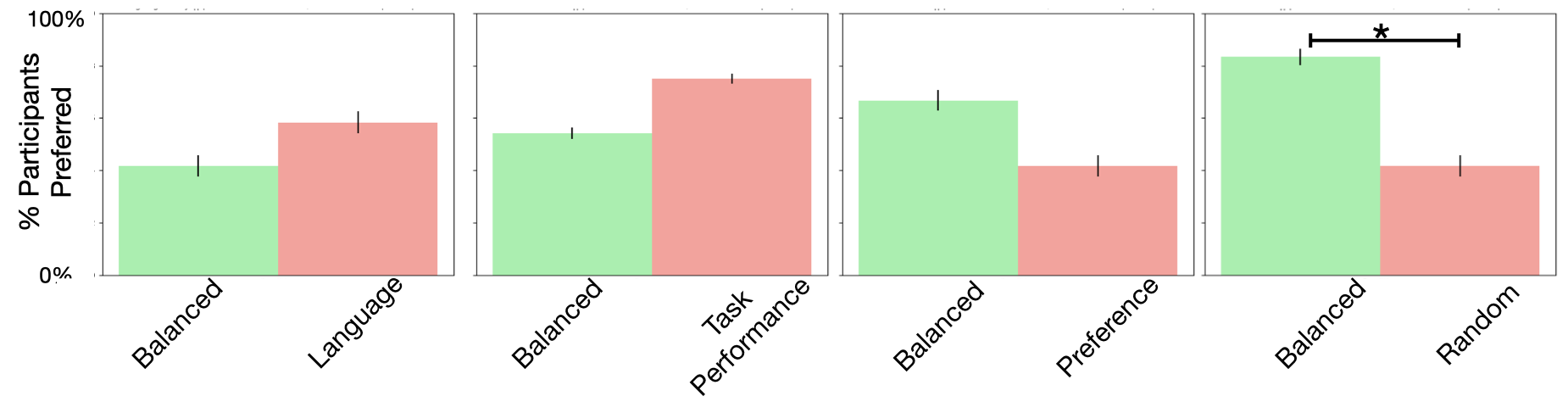}

  \caption{Visualized preference comparison scores across all conditions in the personalization study. Balanced personalization is significantly more preferred than no personalization at all.}
  \label{fig:phase2-pref}
\end{figure*}

\begin{figure*}[!ht]
  \centering
  \includegraphics[width=\textwidth]{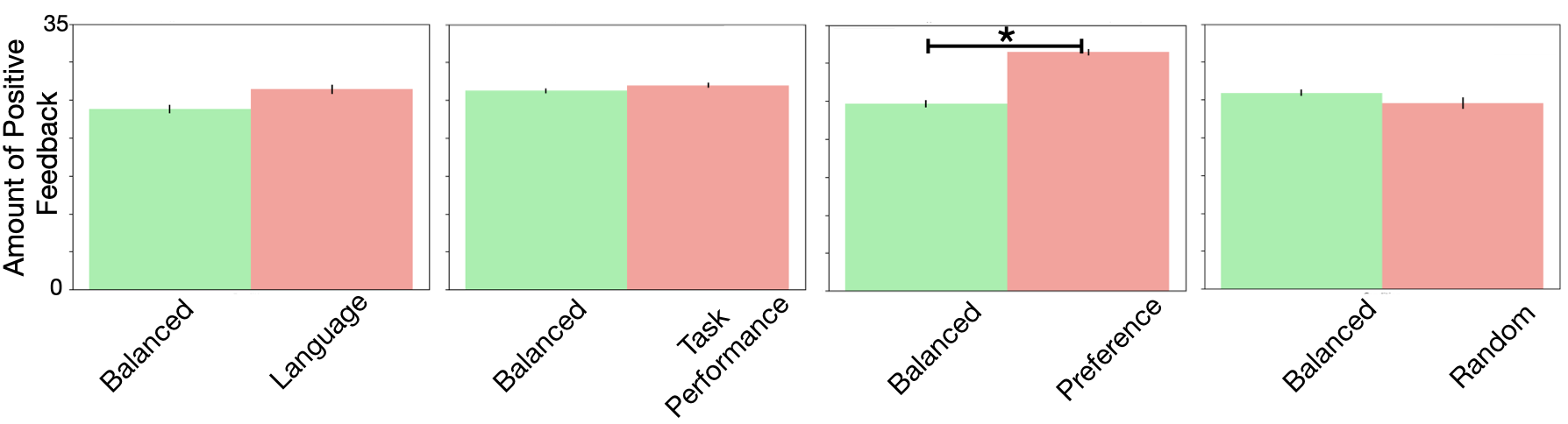}

  \caption{Visualized binary preference feedback across all condition-comparisons in the personalization study. Preference maximization results in significantly more ``Yes'' responses than balanced-personalization.}
  \label{fig:phase2-feedback}
\end{figure*}

\begin{figure*}[!ht]
  \centering
  \includegraphics[width=\textwidth]{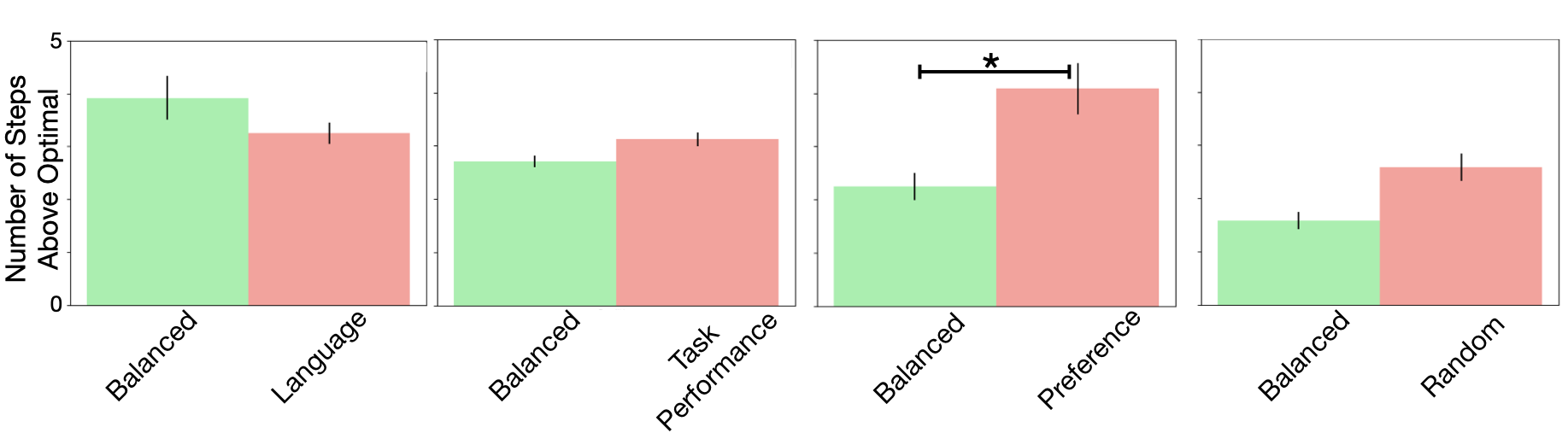}

  \caption{Visualized steps above optimal across all condition-comparisons in the personalization study. Balanced personalization helps participants reach the goal in significantly fewer steps than preference maximization.}
  \label{fig:phase2-steps-above-min}
\end{figure*}

\section{Population Study Statistical Analyses}

An ANOVA for explanation modality rankings yielded a significant difference across baselines ($F(2, 84) = 35.1, p < 0.001$). Data for inappropriate compliance did not pass a Shapiro-Wilk test for normality, and we therefore applied a Friedman's test, which was significant \statcheck{($\chi^2(2)=12.23, p =0.002$).} Data for correct-non-compliance was also not normally distributed, and so we again applied a Friedman's test, which was significant \statcheck{($\chi^2(2)=12.23, p =0.002$).} An ANOVA across explanation modalities for consecutive mistakes was significant \statcheck{($F(2, 74) = 8.0309, p < 0.001$).} Finally, data for consideration time were not normally distributed, and we therefore applied a Friedman's test, which was significant \statcheck{($\chi^2(2)=24.47, p < 0.001$).}

\section{Personalization Study Statistical Analyses}

\subsubsection{Balanced personalization vs. language-only explanations}
A wilcoxon signed rank test for preference rankings did not yield a significant difference across conditions ($W=14, p=0.825$). A Friedman's test over feedback data was not significant ($\chi^2(1)=1.6, p=0.206$). A wilcoxon signed rank test for inappropriate compliance was not significant ($W = 8, p = 0.8688$), nor did a wilcoxon signed rank test for steps above optimal ($W = 37, p = 0.3769$). Finally, an ANOVA across personalization approaches for consideration time was not significant ($F(1, 11) = 4.6718, p = 0.054$).

\subsubsection{Balanced personalization vs. task-performance-based personalization}
A wilcoxon signed rank test for preference rankings did not yield a significant difference across conditions ($W=35, p=0.9585$). A Friedman's test over feedback data was not significant ($\chi^2(1)=2.5789, p=0.108$). 
A wilcoxon signed rank test for inappropriate compliance was not significant ($W = 99, p = 0.1463$), as did a wilcoxon signed rank test for steps above optimal ($W = 97, p = 0.314$).
Finally, an ANOVA across personalization approaches for consideration time was not significant ($F(1, 23) = 0.066, p=0.799$).

\subsubsection{Balanced personalization vs. preference-based personalization}
A wilcoxon signed rank test for preference rankings did not yield a significant difference across conditions ($W=3, p=0.117$). We applied a Friedman's test for feedback data, which revealed a significant difference \statcheck{($\chi^2(1)=5.444, p=0.0196$)}. A wilcoxon signed rank test for inappropriate compliance found significance \statcheck{($W = 21, p = 0.01776$),} as did a wilcoxon signed rank test for steps above optimal \statcheck{($W=31, p=0.03818$)}. 
Finally, an ANOVA across personalization approaches for consideration time was not significant ($F(1, 11) = 0.1634, p=0.694$).

\subsubsection{Balanced personalization vs. random explanations}
A wilcoxon signed rank test for preference rankings did yield a significant difference across conditions \statcheck{($W=4, p=0.0363$)}. A Friedman's test over feedback data was not significant ($\chi^2(1)=0.3173, p=0.317$). 
A wilcoxon signed rank test for inappropriate compliance found significance \statcheck{($W = 25, p = 0.03275$)}, though a wilcoxon signed rank test for steps above optimal did not find significance ($V = 20.5, p = 0.1531$).
Finally, data for consideration time did not pass normality assumptions, and a Friedman's test was not significant ($\chi^2(1)=0.33, p=0.564$).

\subsubsection{Participant Recruitment}
We recognize that our results occasionally include large effects that are not statistically significant, and that such results may become statistically significant with a larger sample population. Our study featured over 100 participants (including pilots), but a power analysis of our preference-ranking results revealed that we would need over 60 participants for a significant effect in the preference-based personalization vs. balanced-personalization comparison. Extrapolating to the rest of our work, we would have required over 240 participants for the study. We leave such a thorough investigation to future work.

\section{Participant Briefing}
\label{app:briefing}
Thank you for participating in our study! Before we get started I need you first to fill out this consent and data release form, please take your time to review it and let me know if you have any questions. 

\textit{<Participant completes consent form>}

Thank you! So today you are going to be helping to guide a self-driving car through a simulated city. The car will handle all of the actual control, and you will be responsible for commanding the car where to go at each intersection in the city. 
You will be given navigational assistance from the self-driving car in the form of on-screen prompts, so the car will tell you which way to go in order to get to the goal as fast as possible. The simulator will pause while the agent thinks about which way to go, and it takes about 4-5 seconds at each intersection for the agent to produce a suggestion. The agent will also present you with short explanations for why it’s giving you a directional suggestion. You can decide at each intersection whether you want to go with what the agent suggests. The car may occasionally not allow you to travel in a direction that seems open. This happens if the car has not yet fully mapped a given turn.  

These explanations can come in three different forms, either written descriptions, decision trees the car uses, or feature importance maps. And for reference, here is an example of what each of those looks like: 

\textit{<Participant is shown example written description reading: ``You should bring lunch to work today because the restaurants in your area will be closed for a holiday.''>}

In the written descriptions, you’ll see a sentence explaining why one choice is better than another. 

\begin{figure*}[!ht]
  \centering
  \includegraphics[width=\textwidth]{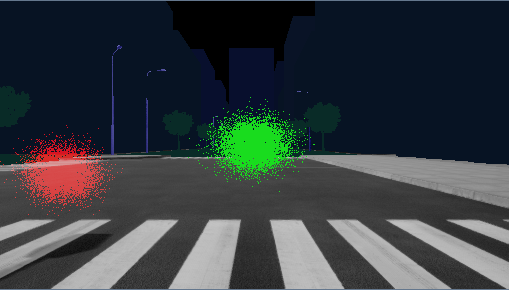}

  \caption{Feature importance map example shown to participants during the instructions phase of the study.}
  \label{fig:right_feature_map}
\end{figure*}

\textit{<Participant is shown an example feature importance map, shown in Figure \ref{fig:right_feature_map}.>}

With the feature importance map, you’ll see a highlighted image with relevant elements of the image highlighted in different colors, such as the outlines of trees and buildings next to the road. The green blob highlights the direction of the best path, whereas the other red blobs highlight the other possible directions that were not chosen due to obstacles or car crashes.  

\begin{figure*}[!ht]
  \centering
  \includegraphics[width=\textwidth]{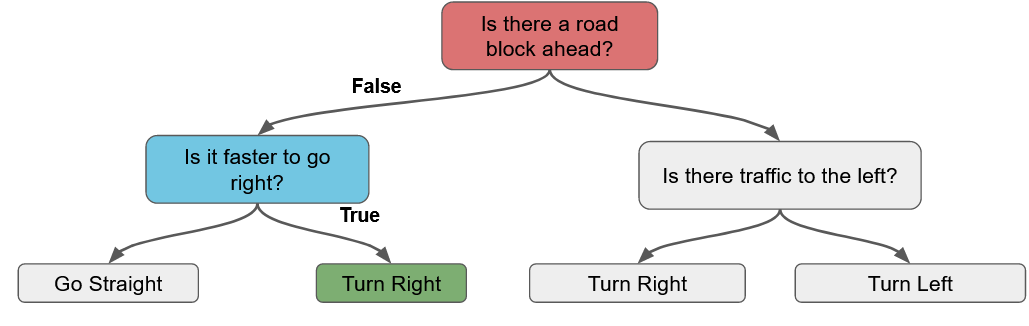}

  \caption{Decision Tree example shown to participants during the instructions phase of the study.}
  \label{fig:right_dt}
\end{figure*}

\textit{<Participant is shown an example decision tree, depicted in Figure \ref{fig:right_dt}.>}

In the decision tree, you’ll see a flowchart with true/false checks that lead to a decision, where a ``true'' check is labeled as true and highlighted in blue, and a ``false'' check is labeled with false and highlighted in red. 

\begin{figure*}[!ht]
  \centering
  \includegraphics[width=0.3\textwidth]{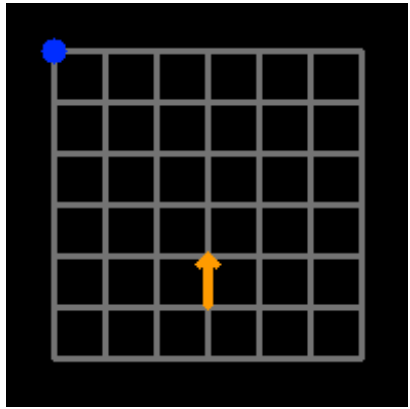}

  \caption{Mini-map example shown to participants during the instructions phase of the study.}
  \label{fig:example_mini_map}
\end{figure*}

\textit{<Participant is shown an example mini-map, given in Figure \ref{fig:example_mini_map}.>}

You will also have access to a mini-map at each intersection, which will look like this. The arrow indicates your current position and heading, and the blue circle is your destination. 

Before we begin, could you please fill out the following pre-study survey for me, and please stop when the survey asks you not to advance further: 

\textit{<Participant takes pre-study surveys>}

Great, thank you! So now we will begin the actual study! As you are helping the car to navigate through the city, you will go through 9 navigation tasks. The first will be a training task for you to become acclimated to working with the car and the self-driving agent, then we’ll do 2 calibration tasks for the agent to learn to accommodate your behaviors. After that, you’ll navigate with one agent for 3 tasks, then we’ll stop to do a couple of surveys. Finally, you’ll resume the task with a new agent for 3 final tasks, and we’ll conclude with a final set of surveys. For each task, the car will reset to a new location in the city, and the goal location will move.  

There are a few important things to bear in mind: 

First, the city is littered with construction projects and car crashes that can block certain routes, and these car crashes and construction projects can move around when the car resets. 

Second, the agent is not perfect and will sometimes make mistakes. The agent is good at estimating obstacles on your shortest path to the goal. However, when it focuses on external factors such as the time of day, the sky, rush hour traffic, the weather, etc., that means the agent is malfunctioning. If this happens, the agent is giving you the wrong direction! In such cases, you should do the opposite of what the agent says, to the extent possible. Even when the explanation is wrong the map is correct. Note that this can happen regardless of the explanation, whether it is language, decision tree or feature map. Remember, if it is looking at external criteria such as time of day, rush hour traffic, the sky, the music on the radio, or the weather, it is wrong.  

Here are examples of when each is wrong.

\textit{<Participant is shown language incorrect language explanation reading: ``You should turn left because the radio is set to NPR.''>}

So we see here the language refers to the radio, which is an external factor so the correct decision is to turn right.

\begin{figure*}[!ht]
  \centering
  \includegraphics[width=\textwidth]{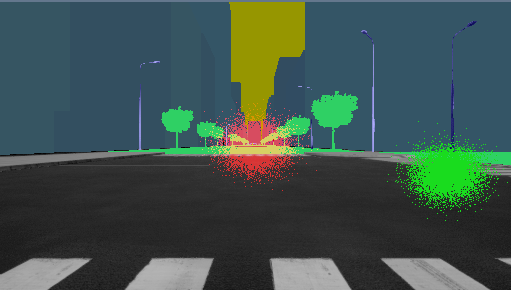}

  \caption{Incorrect feature importance map example shown to participants during the instructions phase of the study.}
  \label{fig:wrong_feature_map}
\end{figure*}

\textit{<Participant is shown an incorrect feature importance map, shown in Figure \ref{fig:wrong_feature_map}.>}

Here we see the feature importance map is looking at the sky, again that is an external factor so you would not follow the agent's suggestion here.

\begin{figure*}[!ht]
  \centering
  \includegraphics[width=\textwidth]{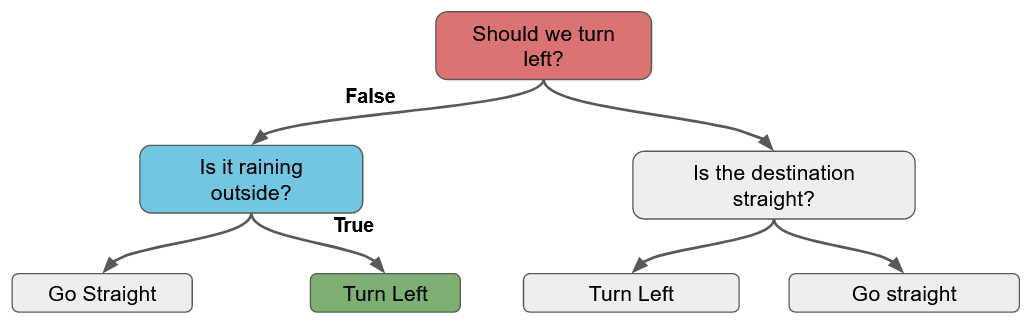}

  \caption{Incorrect decision tree example shown to participants during the instructions phase of the study.}
  \label{fig:wrong_dt}
\end{figure*}

\textit{<Participant is shown an incorrect decision tree, shown in Figure \ref{fig:wrong_dt}.>}

And here we see that the decision tree is considering the weather, which again is an external factor, so you would go right instead of left. Of note, just because this node is incorrect, doesn’t mean the entire tree is wrong, so if you were to see an explanation using the other half of this tree, for example, you could trust that suggestion. But you would not trust it if the decision used an external feature. 

Third, you will be timed during the main body of the task, and we’d like for you to complete each task as quickly as possible. But, the timer will be paused while you make up your mind at each intersection, so you aren’t penalized for taking time to make up your mind on which way you want to go. When you make a choice, please be mindful that you cannot undo it! Once you click “OK”, the car will start to drive on, so be sure you choose the direction you want to go. 

Finally, you have a maximum of 20 total interactions per task. So if you cannot reach the goal within 20 intersections, the task will end and immediately progress to the next one. 


\end{document}